\documentclass[12pt]{iopart}

\usepackage{iopams}  

\usepackage{graphicx}

\newcommand{\be}{\begin{equation}}
\newcommand{\ee}{\end{equation}}
\newcommand{\bea}{\begin{eqnarray}}
\newcommand{\eea}{\end{eqnarray}}
\newcommand{\ben}{\begin{enumerate}}
\newcommand{\een}{\end{enumerate}}
\newcommand{\la}{\left\langle}
\newcommand{\ra}{\right\rangle}
\newcommand{\lc}{\left[}
\newcommand{\rc}{\right]}
\newcommand{\lp}{\left(}
\newcommand{\rp}{\right)}

\newcommand{\dat}{\mathrm{dat}}
\newcommand{\art}{\mathrm{art}} 
\newcommand{\rep}{\mathrm{rep}}
\newcommand{\net}{\mathrm{net}}

\newcommand{\stat}{\mathrm{stat}}

\newcommand{\tot}{\mathrm{tot}}

\newcommand{\NNe}{\mathrm{NN}\lp E_{\nu}\rp}

\newcommand{\draft}[1]{}

\begin{document}


\title{Extraction of the atmospheric neutrino fluxes from 
experimental event rate data}

\author{M. C. Gonzalez-Garcia} 
\address{
  C.N. Yang Institute for Theoretical Physics
State University of New York at Stony Brook
Stony Brook, NY 11794-3840, USA,
and Instituto de F\'\i sica Corpuscular, Universitat de Val\`encia
  -- C.S.I.C.  Edificio Institutos de Paterna, Apt 22085, 46071
  Val\`encia, Spain}
   \ead{concha@insti.physics.sunysb.edu}
\author{ M. Maltoni}
\address{
International Centre for Theoretical Physics, 
Strada Costiera 11, 31014 Trieste, Italy }
\ead{mmaltoni@ictp.trieste.it}
\author{J. Rojo}
\address{
Dept. d'Estructura i Constituents de la Mat\`eria, U. de Barcelona,
Diagonal 647, E-08028 Barcelona, Spain}
\ead{joanrojo@ecm.ub.es}

\begin{abstract}

The precise knowledge of the atmospheric neutrino fluxes
is a key ingredient in the interpretation of the results from 
any atmospheric neutrino experiment. 
In the standard atmospheric neutrino data analysis, these fluxes are 
theoretical inputs obtained from  sophisticated numerical 
calculations.
In this contribution we present an alternative
approach to the determination of the atmospheric neutrino fluxes based 
on the direct  extraction from the experimental data on neutrino event rates.
The extraction is achieved  by means of a combination of artificial neural 
networks as interpolants and 
Monte Carlo methods.


\end{abstract}

\submitto{\JPA}

\section{Introduction}
\label{intro}

One of the most important breakthroughs in particle physics, and
the only solid evidence for physics beyond the Standard Model, 
is the discovery -- following a variety of independent
experiments \cite{neutrev2} -- that neutrinos are massive and 
consequently  can oscillate among their different flavor eigenstates.
The flavor oscillation hypothesis has been supported by an impressive 
wealth of experimental data,  one of the most important pieces of 
evidence coming from atmospheric neutrinos.

Atmospheric neutrinos are originated by the collisions of cosmic rays
with air nuclei in the Earth atmosphere.  These neutrinos are
observed in underground experiments using different
techniques~\cite{sk}. In particular in the last 
ten years, high
precision and large statistics data has been available from the
SuperKamiokande experiment~\cite{sk} which has clearly
established the existence of a deficit in the $\mu$-like atmospheric
events with the expected distance and energy dependence from
$\nu_\mu\rightarrow \nu_\tau$ oscillations with oscillation parameters
$\Delta m^2 \sim 2\times 10^{-3}$ eV$^2$ and $\tan^2\theta=1$.  This
evidence has also been confirmed by other experiments, both
athmosferic such
as MACRO and Soudan 2 and accelerator-based like MINOS and K2K.

The expected number of atmospheric neutrino events depends on a variety
of components: the atmospheric neutrino fluxes,
the neutrino oscillation parameters and the neutrino-nucleus
interaction cross section. Since the main focus of
atmospheric neutrino data interpretation has been the determination
of neutrino oscillation parameters, in the standard analysis  
the remaining components of the event rate computation are inputs 
taken from other sources. In particular, 
the fluxes of atmospheric neutrinos 
are taken from the results of numerical calculations, 
like those of Refs. \cite{honda,bartol}, which make use of a mixture
of primary cosmic ray spectrum measurements, models for the hadronic 
interactions and simulation of particle propagation \cite{fluxrev}.

The attainable accuracy in the independent determination 
of the relevant neutrino oscillation parameters from  non-atmospheric 
neutrino experiments 
makes it possible to attempt an inversion of the strategy in the
atmospheric neutrino analysis: to 
use the oscillation parameters as inputs in the data analysis in order 
to extract from data the atmospheric neutrino fluxes.  

There are several motivations for such direct determination of the
atmospheric neutrino flux from experimental data.  First of all it
would provide a cross-check of the standard flux calculations as well
as of the size of the associated uncertainties (which being mostly
theoretical are difficult to quantify).  Second, a precise knowledge
of atmospheric neutrino flux is of paramount 
importance for high energy neutrino
telescopes \cite{icecubelect}, both because they are the main
background and they are used for detector calibration.  Finally, such
program will quantitatively expand the physics potential of future
atmospheric neutrino experiments.  Technically,
however, this program is challenged by the absence of a generic
parametrization of the energy and angular functional dependence of the
fluxes which is valid in all the range of energies where there is available
neutrino data. 

In this contribution we present the first results on this alternative approach
to the determination of the atmospheric neutrino fluxes: we will
determine these fluxes from experimental data on atmospheric neutrino
event rates, using all available information on neutrino oscillation
parameters and cross-sections.  
The problem of the unknown functional form for the neutrino flux is 
bypassed by the use of neural networks as interpolants, since they
allow us to parametrize the atmospheric neutrino flux 
without having to assume any functional behavior. Additional
details on our approach can be found in \cite{paper}.

Indeed the problem of the {\it deconvolution} of the atmospheric flux from
experimental data on event rates is rather close in spirit to the
determination of parton distribution functions in deep-inelastic
scattering from experimentally measured structure functions
\cite{qcdbook}.  For this reason, in this work we will apply
for the determination of the atmospheric neutrino fluxes a general
strategy originally designed to extract parton distributions in an
unbiased way with faithful estimation of the uncertainties\footnote{
This strategy has also been successfully applied with different
motivations in other contexts like tau lepton decays \cite{tau} and B
meson physics \cite{bmeson}} \cite{f2ns,f2nnp,pdf2}.

\section{General strategy}

\label{genstrat}

The general strategy that will be used to determine the
atmospheric neutrino fluxes was first presented in Ref.~\cite{f2ns}
(see also \cite{nnthesis}). 
It involves two distinct stages in order to go from the data to the
flux parametrization. In the first stage, a 
Monte Carlo sample of replicas of the experimental data on neutrino
event rates  (``artificial data'') is generated. These can be
viewed as a sampling of the probability measure on the space of
physical observables at  the discrete points where data exist. 
In the second stage one uses neural networks to interpolate between
points where data exist. In the present case, this
second stage in turn consists of two sub-steps: the determination of
the atmospheric event rates from the atmospheric flux
in a fast and efficient way, and
the comparison of the event rates thus computed to the data
in order to  tune the best-fit form of input  neural 
flux distribution
(``training of the neural network''). 

In the present analysis we use the latest data on atmospheric
neutrino event rates from the Super Kamiokande Collaboration \cite{sk}.
Higher energy data from neutrino telescopes like
Amanda \cite{amandadata} is not publicly available in a format which
allows for its inclusion in the present analysis and its treatment is 
left for future work.

The latest Super Kamiokande atmospheric neutrino data sample is 
divided in 9 different types of events: contained events in three 
energy ranges, Sub-GeV, Mid-GeV and Multi-GeV 
electron- and muon-like, partially
contained muon-like events  and upgoing stopping and througoing muon events.  
Each of the above types of events is divided in 10 bins in the final
state lepton zenith angle $\phi_l$.
Therefore we have a total of $N_{\dat}= 90$ experimental data points,
which we label as
\be
R_i^{(\exp)}, \qquad i=1,\ldots,N_{\dat} \ .
\ee
Note that each type of atmospheric neutrino event rate is sensitive to
a different region of the neutrino energy spectrum. 
In particular we  note that for energies
larger than $E_{\nu}\sim$ few TeV and smaller than $E_{\nu}\sim0.1$ GeV
there is essentially no information on the atmospheric  flux
 from the available event rates.

The purpose of the artificial data generation is to produce
a Monte Carlo set of `pseudo--data', i.e.
$N_{\rep}$ replicas of the original set of
$N_{\dat}$ data points, $R^{(\art)(k)}_i$
such that the $N_{\rep}$ sets of $N_{\dat}$ points are 
distributed according to an $N_{\dat}$--dimensional 
multi-gaussian distribution around the
original points, with expectation values equal to the central
experimental values, and error and covariance equal to the
corresponding experimental quantities. 
This is achieved by defining 
\be
\label{gen}
R_i^{(\art)(k)}=R_i^{(\exp)}+r_i^{(k)}\sigma_i^{\tot}\ , \quad
k=1,\ldots,N_{\rep} \ ,
\ee
where $N_{\rep}$ is the number of generated replicas
of the experimental data, and
where $r_i^{(k)}$ are univariate gaussian random numbers
with the same correlation matrix as experimental data.

In our case each neural network parametrizes a 
neutrino flux.
Due to available experimental precision, in this work
we will assume the zenith and type dependence of the
flux to be known with some precision and extract from the 
data only its energy dependence. 
That is, if $\mathrm{NN}\lp E_{\nu}\rp$
is the neural network output when the input is
the neutrino energy $E_{\nu}$, then the neural flux
parametrization will be
\be
\Phi^{(\net)}
\lp E_{\nu},c_{\nu},t\rp
=\NNe  \Phi^{\mathrm{(ref)}}\lp E_{\nu},c_{\nu},t\rp \ .
\label{fluxnn}
\ee
$\Phi^{\mathrm{(ref)}}$ is a reference differential flux, which we take
to be the most recent computations of either the Honda
\cite{honda} or the Bartol \cite{bartol} collaborations, 
which have been extended to cover also the high-energy region by 
consistent matching
with the Volkova fluxes.

Now let us describe a fast and efficient technique to
evaluate the atmospheric neutrino event rates
$R_i^{(\net)}$ for an arbitrary input neural
network atmospheric flux $\Phi^{(\net)}\lp E_{\nu},c_{\nu},t\rp$.
As shown in \cite{paper}
it is possible to  write the theoretical predictions
as  a sum of the elements of a bin-integrated flux table
$\Psi^{(\net)}_{ezt}$,
\be
\label{coef}
 R_i^{(\net)}=\sum_{ezt}C^i_{ezt}\Psi^{(\net)}_{ezt} \ ,
\ee
where the coefficients $C^i_{ezt}$, which are the most 
time-consuming ingredient, need only to be precomputed once
before the training, since they do not depend on the
parametrization of the atmospheric neutrino flux.

The determination of the parameters that define the 
neural network, its weights,  is performed by maximum likelihood. 
This procedure, the so-called neural network training, proceeds
by minimizing an error function, which coincides with the $\chi^2$ of 
the experimental points  when compared to their theoretical determination 
obtained using the given set of fluxes:
\be
\label{chi2}
\chi^{2(k)}=\mathrm{min}_{\vec{\xi}}
\Bigg[ \sum_{i}\xi^2_i+
\lp  \sum_{n=1}^{N_{\dat}} \frac{R_n^{\mathrm{(\net)(k)}}
\lc 1+{\displaystyle \sum_i} \pi_i^n\xi_i\rc-R_n^{(\art) (k)}}{
\sigma_n^{\stat}}\rp^2\Bigg] \ .
\ee
The $\chi^2$ has to be minimized with 
respect to $\{ \omega_i \}$, the parameters of the neural network.
The minimization of Eq.~(\ref{chi2}) is performed with the use of 
genetic algorithms \cite{nnthesis}.

Thus at the end of the procedure, we end up with $N_{\rep}$ fluxes, 
with each flux 
$\Phi^{(\net)(k)}$
given by a neural net. The set of $N_{\rep}$ 
fluxes provide our best representation of the  corresponding 
probability density in the space of atmospheric neutrino fluxes:
for example, the mean value of the flux at a given
value of $E_{\nu}$ is found by averaging over the replicas, and the
uncertainty on this value is the variance of the values given 
by the replicas. 

\section{Results and discussion}

\begin{figure}
\centering
  \includegraphics*[scale=0.67]{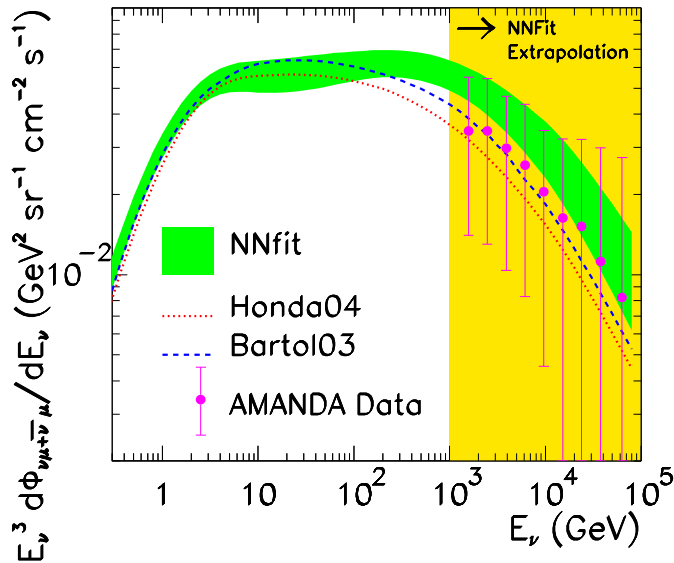}
\caption{Comparison of the $1-\sigma$ uncertainty
band of our parametrization of the atmospheric
neutrino flux with different atmosferic neutrino
flux computations and with the latest
AMANDA data.}
\label{fluxref}       
\end{figure}

\begin{figure}
\centering
  \includegraphics*[scale=0.53]{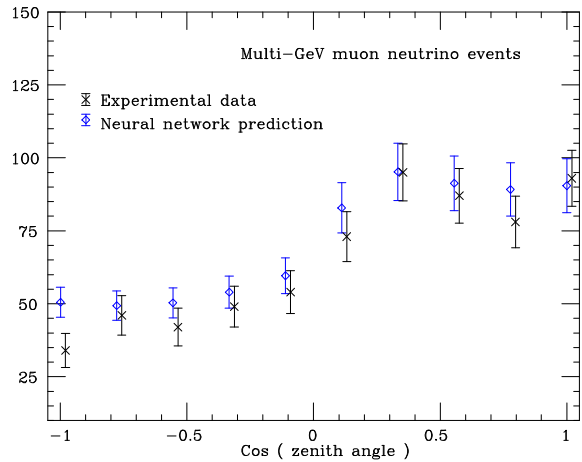}
\caption{Comparison of the neural network results with
experimental data for the Multi-GeV $\nu_{\mu}$
data sample.}
\label{dataplot}       
\end{figure}

 Now we discuss the results of our
determination of the atmosferic neutrino fluxes. 
Note that any functional of the neutrino flux
(like for example event rates and its associated uncertainty)
can be computed taking the appropriate moments over the
constructed probability measure
\be
\la \mathcal{F}\lp \Phi\lp E_{\nu}\rp\rp \ra_{\rep}=
\frac{1}{N_{\rep}}\sum_{k=1}^{N_{\rep}}
 \mathcal{F}\lp \Phi^{(\net)(k)}\lp E_{\nu}\rp\rp \ .
\ee
In Fig. \ref{fluxref} we show our results for the atmosferic
neutrino flux as compared with the Honda and Bartol
computations as well as with some
direct measurements from Amanda \cite{amandadata}. 
Note that above 1 TeV we are in the extrapolation
region.
We show in Fig. 
\ref{dataplot} how our neural network parametrization successfully 
reproduces the features of experimental data. A more detailed
discussion of the results can be found in
\cite{paper}

 The work presented in this contribution
can be extended in several directions. First of all
one could reduce the uncertainty in the large $E_{\nu}$
region by incorporating in the fit atmosferic neutrino
measurements from neutrino detectors like Amanda \cite{amandadata2,
amandadata}. 
Second, we could use at even larger energies 
simple parametrizations of the flux to obtain a better estimation
of backgrounds at neutrino telescopes.
Finally, we could estimate the required statistics that would
be required in forthcoming atmosferic neutrino experiments
in order to determine from data also the zenith angle and
flavor dependence of the neutrino flux.

\vspace{1cm} 
{\bf References}

\vspace{0.5cm}

\providecommand{\newblock}{}

\end{document}